\documentclass[graybox]{svmult}

\usepackage{mathptmx}       
\usepackage{helvet}         
\usepackage{courier}        
\usepackage{type1cm}        
\usepackage{makeidx}         
\usepackage{graphicx}        
\usepackage{multicol}        
\usepackage[bottom]{footmisc}

\makeindex             

\begin{document}

\title*{Astrometric and photometric mass functions of the old
  open cluster Praesepe from the UKIDSS GCS}
\titlerunning{Mass function of Praesepe}
\author{S. Boudreault$^{1,2,\dag}$, N. Lodieu$^{1,2}$, N.  R. Deacon$^{3}$
  and N. C.  Hambly$^{4}$}
\institute{$^{1}$ Instituto de Astrof\'{i}sica de Canarias (IAC),
  C/V\'{i}a L\'{a}ctea s/n, E-38200 La Laguna, Tenerife, Spain; $^{2}$
  Departamento de Astrof\'{i}sica, Universidad de La Laguna (ULL),
  E-38205 La Laguna, Tenerife, Spain; $^{3}$ Max-Planck-Institute
  f\"{u}r Astronomie, K\"{o}nigstuhl 17, 69117, Heidelberg, Germany;
  $^{4}$ Scottish Universities\~ Physics Alliance (SUPA), Institute
  for Astronomy, School of Physics \& Astronomy, University of
  Edinburgh, Royal Observatory, Blackford Hill, Edinburgh EH9 3HJ, UK;
  $^{\dag}$\email{szb@iac.es}}
\maketitle

\abstract*{Here we present the results of a wide-field ($\sim$36 sq.
  deg.)  near-infrared ($ZYJHK$) survey of the Praesepe cluster using
  the Data Release 9 (DR9) of the UKIRT Infrared Deep Sky Survey
  (UKIDSS) Galactic Clusters Survey (GCS). We selected cluster
  candidates of Praesepe based on astrometry and photometry. With our
  candidate list, we have obtained the luminosity function of Praesepe
  in the $Z$ and $J$ bands, and we have derived the mass function (MF)
  of Praesepe from 0.6 down to 0.072\,M$_\odot$. Moreover, we have
  estimated the binarity of the Praesepe members in the
  0.5-0.1\,M$_\odot$ mass range and as well as their variability.}

Here we presented the results of a wide field, near--infrared study of
the Praesepe cluster using the DR9 of the UKIRT Infrared Deep Sky
Survey Galactic Clusters Survey. We performed an astrometric and
photometric selection of 1,116 cluster candidates out of the 218,141
point sources detected towards Praesepe.

Possible sources of contamination include Galactic disk late-type and
giant stars and unresolved galaxies.  We estimate a contamination rate
of 11.9\,\% above 0.4\,M$_\odot$, 9.8\,\% in the mass range
0.15--0.4\,M$_\odot$, and 23.8,\% below 0.15\,M$_\odot$.

We investigated the binary frequency of Praesepe using the photometry
and colours from our cluster candidates. We observe a binary fraction
similar to the simulation of Bate (2012) between 0.07--0.1\,M$_\odot$,
$\sim$1.5$\sigma$ difference in the 0.2--0.45\,M$_\odot$ mass
interval, and significantly lower by more than 3$\sigma$ for the mass
range 0.1--0.2\,M$_\odot$. On the other hand, the binary fraction from
Pinfield et al. (2003) are higher than our values and those of Bate
(2012). We note that two other works focusing on field low-mass stars
have also derived binary fractions lower than Bate (2012).

We also studied the variability of the Praesepe candidates using the
two $K$--band epochs provided by the GCS. We identified seven
candidate variables, including one in the substellar regime.

We derived the luminosity function of Praesepe in $Z$ and $J$--band
here. We observed that the peak of the $J$--band luminosity function
is one magnitude brighter than the one reported by
Boudreault et al. (2010).

\begin{figure}[b]
\includegraphics[scale=0.7]{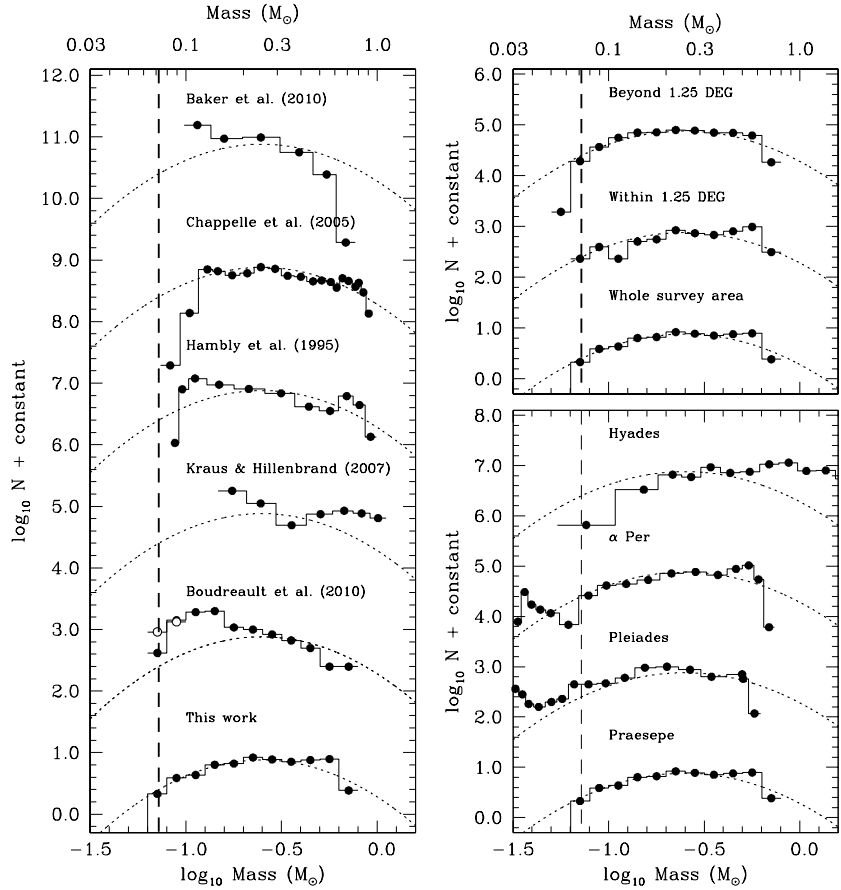}
\caption{For all panels, the dotted curved lines is the system
  Galactic field star MF fit from Chabrier (2005), the vertical dashed
  line, and the normalization of all the MFs at $\sim$0.3\,$M_{\odot}$
  are the same as in. We normalised all the MFs to the log-normal fit
  of Chabrier (2005) at $\sim$0.3\,$M_{\odot }$ ($\log M \sim -
  0.5$).}
\label{figure1}
\end{figure}

Finally, we determined the mass function (MF) of Praesepe, which
differs from previous studies: while previous MFs showed an increase
from 0.6 to 0.1\,M$_\odot$, our MF shows a decrease. We looked at the
MF of Praesepe at two different regions of the cluster, i.e.~within
and beyond 1.25$^{\circ}$, and we observed that both regions show an
MF which decreases to lower masses. We compared our MF of Praesepe in
the mass range 0.072--0.6\,M$_\odot$ with the ones of the Hyades, the
Pleiades and $\alpha$~Per. We conclude that our MF of Praesepe is most
similar to the MF of $\alpha$~Per although they are respectively of
$\sim$85 and $\sim$600\,Myr. Even though of similar age, the Praesepe
remains different than the Hyades, with a decrease in the MF of only
$\sim$0.2\,dex from 0.6 down to 0.1\,$M_{\odot}$, compared to
$\sim$1\,dex for the Hyades. All MFs are presented in
Fig.~\ref{figure1}.

\begin{acknowledgement}
  SB and NL are funded by national program AYA2010-19136 (Principal
  Investigator is NL) funded by the Spanish ministry of science and
  innovation. NL is a Ram\'on y Cajal fellow at the IAC (program
  number 08-303-01-02).
\end{acknowledgement}

\end{document}